\documentclass[epj]{svjour}

\usepackage[dvips]{graphicx}
\usepackage{amssymb}

\begin{document}
\title{Frustrated Rotations in Nematic Monolayers}
\author{Y. Tsori \and P.-G. de Gennes
}                     

\institute{Physique de la Mati\`{e}re Condens\'{e}e, Coll\`{e}ge de 
France,
Paris, France}
\date{Received: date / Revised version: date}
%
\abstract{
Tabe and Yokoyama found recently that the optical axis in a chiral
monolayer of a ferronematic rotates when water evaporates from the
bath: the chiral molecules act as propellers. When the axis is
blocked at the lateral walls of the trough, the accumulated
rotation inside creates huge splays and bends. We discuss the
relaxation of these tensions, assuming that a single dust particle
nucleates disclination pairs. For the simplest geometry, we then
predict a long delay time followed by a non-periodic sequence of
``bursts''. These ideas are checked by numerical simulations.
\PACS{
      {61.30.Hn}{Surface phenomena: alignment, anchoring, anchoring 
transitions, surface-induced layering, surface-induced ordering, wetting, 
prewetting transitions, and wetting transitions}   \and
      {61.30.Jf}{Defects in liquid crystals}
     } 
} 
\maketitle

\section{Introduction}

The chiral compound R(OPOB)
\begin{figure}[h!]
\begin{center}
\includegraphics[scale=0.4,bb=40 650 575 750,clip]{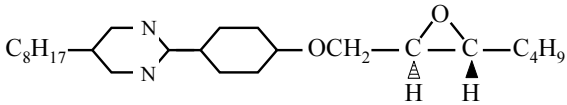}
\end{center}
\caption{Chemical structure of R(OPOB).}
\end{figure}
forms a bulk ferroelectric liquid crystal. But it can also be
deposited as a monolayer on water (or on water glycerol mixtures,
which slow down the dynamics). Tabe and Yokoyama observed these
layers optically and found that the optical axis of this
two-dimensional nematic was rotating continuously in the $x$-$y$
plane \cite{TY}. They interpreted this remarkable result in terms
of an irreversible process: water evaporation from the supporting
liquid is responsible for the molecular precession. They wrote
down the basic Onsager relations required to describe the effect.
They consist of

a) Two fluxes: the rotation rate $\Omega$ and the mass flux of
water evaporation $J_w$.

b) Two forces: the torque $\tau$ (around a vertical axis) applied
on the nematic molecules by external fields or by local
distortions, and the chemical potential difference $\Delta\mu$
between water in the liquid and water in the gas \cite{remark1}.

The entropy source per unit area of the monolayer is then
\begin{eqnarray}
T\dot{S}=\tau\Omega+\Delta\mu J_w
\end{eqnarray}
and the Onsager relations are
\begin{eqnarray}
\Omega &=&\gamma_1^{-1}\tau+b\Delta\mu\label{on1}\\
J_w &=&b\tau +P\Delta\mu\label{on2}
\end{eqnarray}
In Eq. (\ref{on1}),
$\gamma_1$ is a classical friction coefficient for nematic
rotations \cite{pggp} and $P$ is related to
the
permeability of
the air layer above the Langmuir trough. The interesting coupling
coefficient is $b$.

Ref. \cite{TY} shows that in simple conditions ($\Delta\mu$ is
fixed and $\tau\sim 0$), $\Omega$ is indeed proportional to
$\Delta\mu$, and that the experimental values of $b$ are of the
right order of magnitude (typically the rotation periods
$2\pi/\Omega$ are a few seconds).

We are interested here in one particular feature - the angle
$\phi$ (labelling the direction of the optical axis in the
horizontal plane) is forced by water evaporation to rotate in the
monolayer. But at the lateral edges of the trough, the direction
is expected to be anchored at the walls \cite{pggp}. This
frustrated situation forces the internal elastic energy
(proportional to $|\nabla\phi|^2$) to relax by successive
``bursts''. Our aim is to present a simple description of these
bursts.

\section{Weak and strong anchoring}

The basic equation for the director angle $\phi(x,y,t)$ in the
monolayer can be derived from Eq. (\ref{on1}), incorporating
elastic torques into $\tau$, and is
\begin{equation}\label{diff_eq}
\frac{\partial\phi}{\partial
t}=\frac{K}{\gamma_1}\nabla^2\phi+\Omega
\end{equation}
where $\Omega$ is fixed because $\Delta\mu$ is imposed. We have
assumed that the splay and bend elastic constants are equal,
$K_1=K_2=K$. It will be convenient to define a diffusion
coefficient
\begin{equation}
D=\frac{K}{\gamma_1}
\end{equation}
which we expect to be of order $10^{-6}$-$10^{-7}$ cm$^2$/sec,
depending on the water/glycerol fraction.

Frustration originates from the anchoring at the lateral
boundaries of the trough. We expect to have an energy per unit
length at the boundary of the order $-U\cos(\phi-\beta)$, where
$\beta$ define a preferred orientation (e.g. normal to the walls)
\cite{remark2}. When expressed per molecule, the anchoring energy
$U$ is small compared to the nematic coupling (maybe 10 times
smaller), but it is very important. If we have a gradient
$\nabla\phi |_n$ normal to the wall, anchoring ruptures whenever
the torque $K\nabla\phi |_n$ becomes larger than $U$. This
corresponds to a critical value of the gradient
\begin{equation}\label{rupture}
\kappa_{\rm surf}=\nabla\phi |_n=U/K\sim 1/10 a
\end{equation}
where $a\sim 1$nm is a molecular size. Thus we see that
$\kappa_s\sim 10^6$ cm$^{-1}$. Here we shall be mostly interested
in the limit where the gradient $|\nabla\phi|$ is smaller than
$\kappa_{\rm surf}$, i.e., the strong anchoring limit. Then
frustration effects in the interior of the monolayer are
important.

\section{The role of the dust particle}
\begin{figure}[h!]
\begin{center}
\includegraphics[scale=0.9,bb=190 560 425 705,clip]{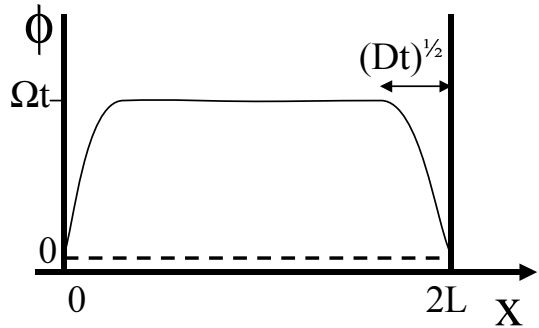}
\end{center}
\caption{A nematic monolayer confined by two walls at $x=0$ and
$x=2L$, with boundary conditions $\phi=0$ at both ends. At $t=0$,
$\phi=0$ everywhere, dashed line. After
time $t$ has passed and under the influence of Eq. (\ref{diff_eq}) the
profile
evolves, solid line.}
\end{figure}

We shall focus our attention on a simple geometry (see Fig. 2),
where the monolayer is limited by two parallel walls separated by
distance $2L$. The boundary conditions are that $\phi=0$ (modulus
$2\pi$) on each wall \cite{remark3}. Assume, for instance, that at
time $t=0$ we have $\phi=0$ everywhere (Fig. 2, dashed line). At a
later time $t$ smaller than $L^2/D$ the angle $\phi$ has rotated
to $\phi=\Omega t$ in a central region, but it drops to zero in a
diffusion region of size $(Dt)^{1/2}$ near each wall (Fig. 2,
solid line). In these regions we have a strong gradient
\begin{equation}\label{str_grd}
\frac{\partial\phi}{\partial x}= c\frac{\Omega
t}{(Dt)^{1/2}}=c\Omega\left(\frac{t}{D}\right)^{1/2}
\end{equation}
where the constant above, $c=2/\sqrt{\pi}$, is taken to be unity in
subsequent calculations \cite{c}.

When this gradient reaches some critical value $\kappa$, a
``burst'' may occur. But we have to be careful: this burst is a
nucleation process for a new structure. Nucleation in nature is
not controlled by the ideal (homogeneous nucleation) threshold. It
is catalysed by a dust particle, a cosmic ray, or some other
perturbation. A realistic formulation requires that we put a
point-like ``dust particle'' somewhere in the trough. We assume
that a burst will occur when the magnitude of the gradient at this
point $|\nabla\phi|$ reaches a critical value $\kappa$. At all
other points $|\nabla\phi|$ may possibly exceed $\kappa$.

Thus we start our discussion by looking at Fig. 3 and
imposing that one dust particle is located at some point $x=d$ and
$y=0$. We impose that $d\ll \sqrt{Dt}\ll L$ to be in the region
described by Eq. (\ref{str_grd}). When the local gradient
$\partial\phi/\partial x$ located at the point $x$ reaches the
critical value $\kappa$, a new scenario must start.

\section{The first burst}
\begin{figure}[h!]
\begin{center}
\includegraphics[scale=0.6,bb=115 510 560 765,clip]{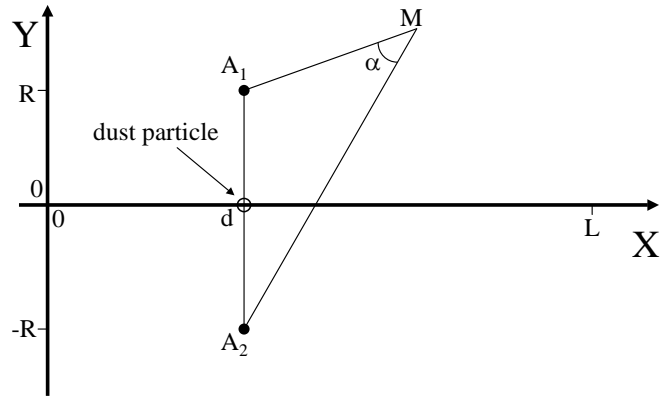}
\end{center}
\caption{Nematic monolayer confined between $x=0$ and $x=L$ under
the influence of driven rotations. The dust particle is at $x=d$
and $y=0$. When elastic stresses are too large, two point
disclinations at $A_1$ and $A_2$ appear.}
\end{figure}

The first breaking event occurs when $\partial\phi/\partial x$
described by Eq. (\ref{str_grd}) reaches the critical value
$\kappa$. This corresponds to a time $t_1=\kappa^2 D/\Omega^2$. We
think of $\kappa$ values of order $10^5$ cm$^{-1}$ (comparable to
the inverse size of the dust particle). Thus $t_1$ is of order $1$
hour and the corresponding diffusion length is $(Dt_1)^{1/2}\sim
1$ mm. Our solution after the burst implies a pair of
disclinations $A_1$ and $A_2$ located at $x=d$ and $y=\pm R(t)$,
see Fig. 3. The initial value of $R$ is comparable to the dust
particle size. The $\phi$ field is then described by
\begin{equation}\label{phi0_phi1}
\phi=x\frac{\partial\phi(t)}{\partial x} +\phi_1=\phi_0+\phi_1
\end{equation}
where the first term corresponds to Eq. (\ref{str_grd}) and $\phi_1$ 
is due to the disclination, and is given simply by
\begin{equation}\label{alpha}
\phi_1(M)=\alpha
\end{equation}
\begin{figure}[h!]
\begin{center}
\includegraphics[scale=0.55,bb=95 220 530 640,clip]{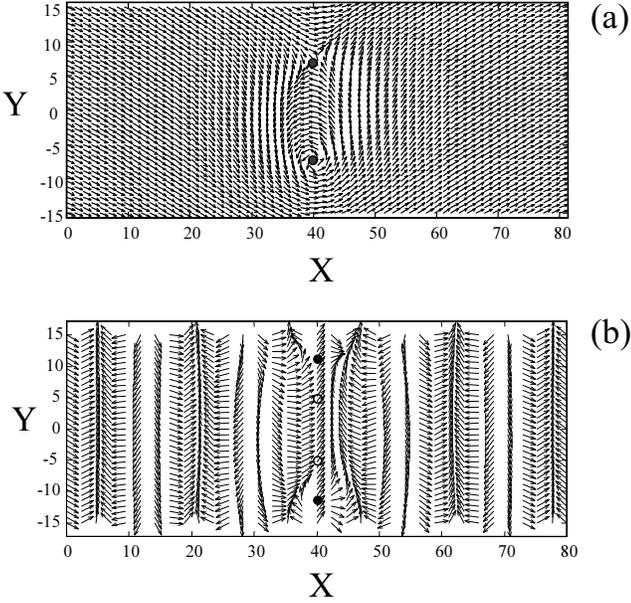}
\end{center}
\caption{(a) Ferro-nematic vector in the $x$-$y$ plane from $\phi_1$
of the
pair disclination of Fig. 3. The two point defects are marked with
filled circles. (b)
orientation vector corresponding to the field $\phi=\phi_0+\phi_1$
given by
Eqs. (\ref{phi0_phi1}) and (\ref{alpha}). Empty and filled circles
mark the
initial and final location of the point defects in the simulation.
Lengths are scaled
by $a=10^{-7}$ cm.}
\end{figure}

Here $\alpha$ is the angle $\widehat{A_1MA_2}$ defined as in Fig.
3, and $M\equiv (x,y)$ is the observation point. $\phi_1$
satisfies the equation $\nabla^2\phi_1=0$, as first noticed by F.
C. Frank (see Fig. 3).  We assume that the distance $2R=A_1A_2$ is
small compared to the wall distance, $R\ll d$ and therefore the
boundary conditions at the wall are satisfied without
including images. Since $\nabla \phi_1$ is opposed in direction to
the unperturbed gradient $\nabla\phi_0$, the disclination pairs
reduces the elastic energy $F_{\rm el}$
\begin{eqnarray}
F_{\rm el}&=&\frac12 K\int \left(\nabla\phi\right)^2~{\rm d}x{\rm
d}y=E_0+E_1+E_2
\end{eqnarray}
where $E_0$ is the unperturbed form, while
\begin{eqnarray}
E_1&=&\int K\nabla\phi_0\nabla\phi_1{\rm d}x{\rm
d}y
\end{eqnarray}
can be reduced to a line integral along $A_1A_2$,
\begin{eqnarray}
E_1=\int_{A_1}^{A_2}K\nabla\phi_0[\phi_1]~{\rm
d}l=-K|\nabla\phi_0|~4\pi
R\label{E1}
\end{eqnarray}
where we introduced a cut along the ($A_1$, $A_2$) segment, and the
discontinuity of $\phi_1$ is $[\phi_1]=2\pi$. Finally, $E_2$ is
the energy of the
disclination pair in an undistorted medium
\begin{eqnarray}\label{E2}
E_2&=&\frac12 K\int \left(\nabla\phi_1\right)^2{\rm d}x{\rm
d}y=
2\pi K \ln\left(\frac{R}{a}\right)
\end{eqnarray}
Here $a$ is proportional to the core radius of the disclination.

The force acting on $A_1$ has an attractive component derived from Eq.
(\ref{E2}) and a repulsive component derived from (\ref{E1})
\begin{eqnarray}\label{f1}
f_1=-\frac12\frac{\partial}{\partial R}(E_1+E_2)=2\pi
K\left(|\nabla\phi_0|-\frac{1}{2R}\right)
\end{eqnarray}
At the moment of burst $|\nabla\phi_0|$ has its threshold value
$\kappa$. The regime of interest is $R\gg \kappa^{-1}$, and the force
is simply $f_1=2\pi \kappa K$. As is known for nematics \cite{pggp}, a
disclination line moving in a constant force has a (nearly) constant
velocity $v$
\begin{eqnarray}\label{v}
v=\dot{R}=f_1/\zeta
\end{eqnarray}
where $\zeta$ is a friction coefficient due to the rotations induced
by the moving line
\begin{eqnarray}
\zeta=\pi\gamma_1\ln\left(\frac{L}{a}\right)
\end{eqnarray}
\begin{figure}[h!]
\begin{center}
\includegraphics[scale=0.6,bb=90 135 510 775,clip]{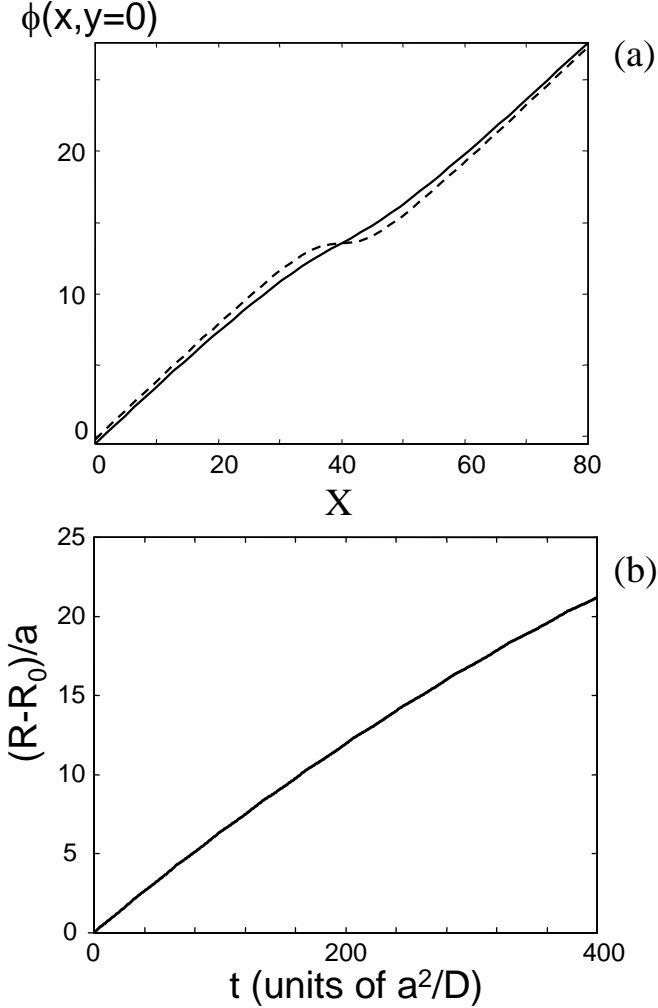}
\end{center}
\caption{(a) dashed line is a plot of $\phi$ as a function of $x$ just
after the
first burst. The  dust particle is at $x=40$ ($x$-axis scaled by
$a=10^{-7}$
cm). The gradient $\phi'(x)$ is reduced
near the particle. Solid line shows the profile at some later time.
(b) plot of the
defect location $R(t)$ as a function of time. The near-linear behavior
is in accord
with Eqs. (\ref{f1}) and (\ref{v}). We used $D=10^{-6}$ cm$^2$/sec and
$\kappa=10^6$ cm$^{-1}$.}
\end{figure}

We conducted a numerical study where $\partial\phi/\partial
t=D\nabla^2\phi$ was solved with different but convenient boundary
conditions, namely $\phi=0$ at the left wall and $\phi= \Omega t$
at the right wall. The dust particle was assumed to be half-way
between the walls. Figure 4 (b) shows the director in the $x$-$y$
plane at some time after the first burst, i.e. after a
disclination pair is created. It consist of a sum of the uniformly
changing field $\phi_0$ and the disclination given by $\phi_1$.
Filled circles mark the defect cores while empty circles mark the
cores just after the burst.

Figure 5 (a) is a plot of $\phi$ as a function of lateral coordinate
$x$, for $y=0$
in Fig. 4 (b). The solid line corresponds to Fig. 4 (b) while the
dashed line is just
after the burst event. Clearly, the gradient $\phi'(x)$ is lowered
near the dust
particle ($x=40$). In Fig. 5 (b) we show the time evolution of the
inter-pair
distance $R(t)$. In the simulation we have $R\kappa\gg 1$ and the
nearly linear
dependence of $R$ on $t$ is evident.

\section{The sequence of bursts}

We return now to the original problem, with an unperturbed
gradient described by Eq. (\ref{str_grd}). It is easy to see that
just after the first burst at $t=t_1=\kappa^2D/\Omega^2$ (and ignoring
a factor of $1/c^2\simeq 0.8$ in Eq. (\ref{str_grd})), the
correction to the gradient near the dust particle is $\partial
\phi_1/\partial x=-2/R$, so that the gradient of $\phi$ in the
$x$-direction is reduced. After the burst, $\phi$ increases with
time. The second burst appears when the gradient becomes again
equal to $\kappa$,
\begin{eqnarray}\label{t2}
\kappa=\frac{\Omega
t_2}{(Dt_2)^{1/2}}-\frac{2}{\dot{R}}\frac{1}{t_2-t_1}
\end{eqnarray}
We introduce the notation
\begin{eqnarray}
\theta_n&=&(\Omega t_n)^{1/2}\\
\lambda&=&(D/\Omega)^{1/2}
\end{eqnarray}
With this notation we find that $\theta_1=\kappa\lambda\simeq
10^2$ is a rather large number. Using $R=\dot{R}t=D\kappa t$ and
$\Omega\lambda/D\kappa=1/\kappa\lambda$, Eq. (\ref{t2}) can be
rewritten in the form
\begin{eqnarray}\label{eq_t2}
\theta_2=\kappa\lambda+\frac{1}{\kappa\lambda(\theta_2^2-
\theta_1^2)}
\end{eqnarray}
Analysis of this equations reveals that $\theta_2$ is rather close to
$\theta_1$:
treating $\kappa\lambda$ as a large number one can obtain that
\begin{eqnarray}\label{th2}
\theta_2\simeq \kappa\lambda+\frac{1}{\sqrt{2}\kappa\lambda}
\end{eqnarray}
The second term on the right hand side is very small compared to the 
first one. 
What is the time difference between the first and second bursts?
If $\Omega=1$ sec$^{-1}$ then $t_2-t_1=1.4$ sec. Thus the bursts
start to be frequent once we wait more than the long time $t_1$.

Using the assumption $\kappa\gg 1/R$ (meaning that the defects
$\phi_1$ interact mainly with $\phi_0$ but not with themselves),
we get for the n'th burst event
\begin{eqnarray}\label{eq_tn}
\kappa=\frac{\Omega
t_n}{(Dt_n)^{1/2}}-\frac{2}{\dot{R}}\left(\frac{1}{t_n-t_{n-1}}+\frac{
1}
{t_n-t_{n-2}}+\right.\nonumber\\
\left.~.~.~.~+\frac{1}{t_n-t_1}\right)
\end{eqnarray}
Or, written differently,
\begin{eqnarray}
\theta_n=\kappa
\lambda+\frac{1}{\kappa\lambda}\sum_1^{n-1}\frac{1}{\theta_
n^2-\theta_m^2}
\end{eqnarray}
The solution for $\theta_2$ obtained above [Eq. (\ref{th2})] suggests 
that the 
sequence of bursts has the form
$\theta_n=\kappa\lambda+An^\alpha$, with $\kappa\lambda\gg 
An^\alpha$.
We put this form of $\theta_n$ and approximate the sum by an integral,
\begin{eqnarray}
An^\alpha\simeq 
\frac{1}{\kappa\lambda}\int_1^{n-1}\frac{1}{2A\kappa\lambda}\frac{dm
}{
n^\alpha-m^\alpha}
\end{eqnarray}
We change variable from $m$ to $p=m/n$:
\begin{eqnarray}
An^\alpha= \frac{n}{2A(\kappa\lambda)^2 
n^\alpha}\int_{1/n}^{1-1/n}\frac{dp}{1-p^\alpha}
\end{eqnarray}
This equation is satisfied if $\alpha=1/2$. In this case the integral 
on the right hand side is equal to $\frac12 
\ln(n-1)+\frac12\ln((2n+1)/(n+1))$, and can be approximated by unity 
because of the logarithmic dependence on $n$. In this approximation we 
find that $A=\frac{1}{\sqrt{2}}\frac{1}{\kappa\lambda}$. In summary, 
the burst sequence is given by
\begin{eqnarray}
\theta_n&=&\kappa\lambda+\frac{1}{\sqrt{2}}\frac{1}{\kappa\lambda} 
n^{1/2}
\end{eqnarray}
The time difference between two successive bursts increases as 
$n^{1/2}$,
\begin{eqnarray}
t_{n+1}-t_n\simeq \frac{1}{\Omega}\sqrt{\frac{n}{2}}
\end{eqnarray}
and is of the order of few seconds.

\section{Conclusion}

We have studied a ferro-nematic monolayer under the influence of
water evaporation. This evaporation introduces an external torque
tending to rotate the individual molecules. Frustration should
occur because the molecules at the edges of the Langmuir trough
are anchored to the walls.

For a simple geometry, we show that a dust particle or some other
disturbance can initiate a pair of disclinations, thus relieving
the stress near the particle. The distance between the two
disclination points increases linearly with time, until the local
strains near the particle again reach their threshold. At this
time two more disclinations are born, also moving away from each
other, and the process continues.

What would happen in a more realistic geometry, for instance with
a circular trough, and anchoring condition which imposes a
director normal to the walls?
In this case, the ideal starting configuration at $t=0$ would have
a disclination point at the center of the trough. When we switch
on the evaporation process at $t=0$, the alignment around this
point would rotate uniformly in time:
$\phi(x,y,t)=\phi(x,y,0)+\Omega t$. But near the walls, in a zone
of width $(Dt)^{1/2}$, we would find strong gradients, and dust
particles could initiate certain ``ladders of bursts''.

There are few practical complications to be expected: \\
\noindent a) The initial state will usually involve many
disclination points rather than one. We still expect a uniform
rotation of the pattern in the central region - this probably
corresponds to the experiment of Ref. \cite{TY}. One could remove
the disclination from the central region by preparing the sample
under a horizonal magnetic field $H$, and then switching off the
field at $t=0$. \\ \noindent b) Our description of the
nemato-hydrodynamics was simplistic \cite{pggp}. With a fuller
description we might expect certain backflow effects - the dust
particles might move in the $x$-$y$ plane. \\ \noindent c) The
width $(Dt_1)^{1/2}$ of the diffusion layer (at the onset of the
ladders) is expected to be rather small, about a millimeter. In
this region, the monolayer need not be flat, as a meniscus will be
created near the walls.

A final remark: there is a superficial analogy between rotations
due to evaporation, which we discuss here, and the rotations
induced in a nematic by a rotating magnetic field \cite{pggp}. 
But there is a deep difference. When the field is present, the 
rotational 
symmetry is broken, and the defects generated by frustration are {\it 
domain 
walls}. But in our case, rotational symmetry is preserved and the 
relevant defcts 
are {\it disclination lines}.

\section{Appendix- burst sequence in one dimension}
We now discuss the series of burst events in the one-dimensional
case, namely $\phi$ depends on one variable only, $x$. The time
$t_1$ where the first event occurs is given by
\begin{eqnarray}
\kappa=\frac{\Omega t_1}{(Dt_1)^{1/2}}
\end{eqnarray}
In our prescription, the defect is created in a ``delta function''
manner, i.e. one
molecule changes its orientation. After some time, this single
molecule affects
its neighborhood in a diffusive manner. Hence we write that the slope
is reduced
by a value
$\sqrt{\pi/D(t-t_1)}\exp\left(-(x-d)^2/\sqrt{2D(t-t_1)}\right)$, where
$d$ is the particle location. At the second burst event, the total
slope
$\phi'(d)$ is again equal to the threshold value $\kappa$, and hence
we have
at $t_2$
\begin{eqnarray}
\kappa=\frac{\Omega
t_2}{(Dt_2)^{1/2}}-\frac{\sqrt{\pi}}{\sqrt{D(t_2-t_1)}}
\end{eqnarray}
Similarly, the n'th event is given by
\begin{eqnarray}
\kappa=\frac{\Omega t_n}{(Dt_n)^{1/2}}-\sqrt{\frac{\pi}{D}}
\left(\frac{1}{\sqrt{t_n-t_1}}+\frac{1}{\sqrt{t_n-t_2}}+\right.
\nonumber\\
\left..~.~.~+
\frac{1}{\sqrt{t_n-t_{n-1}}}\right)
\end{eqnarray}
Using $\theta_n=(\Omega t_n)^{1/2}$ and $\lambda=(D/\Omega)^{1/2}$,
we rewrite the last equation as
\begin{eqnarray}\label{app_thn}
\theta_n=\lambda\kappa+\pi^{1/2}\sum_1^{n-1}\frac{1}{\sqrt{\theta_n^
2-\theta_m^2}}
\end{eqnarray}
We continue to solve this relation assuming that
$\theta_n=An^\alpha+\lambda\kappa$, and that on the right hand side of
Eq.
(\ref{app_thn}), $\lambda\kappa$ is larger than the sum.  We also
approximate the sum by an integral, and find that
\begin{eqnarray}
An^\alpha=\frac{\pi^{1/2}}{\sqrt{2\lambda\kappa
An^\alpha}}\int_0^n\frac{dm}{\sqrt{1-\left(\frac{m}{n}\right)^\alpha}}
\end{eqnarray}
This equation is satisfied when $\alpha=2/3$: in this case the
variable in the
integral can be changed from $m$ to $p=nm$, and the integral becomes
equal to
$3\pi/4$. In summary,
\begin{eqnarray}
\theta_n&=&An^\alpha+\lambda\kappa\\
n&=&2/3\nonumber\\
A&=&\frac18\pi\left(\frac{3}{4}\right)^{3/2}\frac{1}{(\lambda\kappa)^3
}
\nonumber
\end{eqnarray}

The above sequence of events could be relevant for a one
dimensional superconductor rod. If the rod is thin, in each
cross-section the phase is constant. When a current is driven into
the system, this phase changes continuously along the rod. When
the phase gradient is too large, a similar defect should be
created, nucleating most probably around a mechanical ''weak
spot''. The detailed nucleation process was discussed long ago by
Langer and Ambegaokar \cite{ambeg}.


\begin{thebibliography}{}

\bibitem{TY} Y. Tabe and H. Yokoyama, Nature Materials
\textbf{2}, (2003) 806.

\bibitem{remark1} For convenience we define $\Delta\mu$ for unit
mass of the water rather than per water molecule.

\bibitem{pggp} P.-G. de Gennes and J. Prost, \textit{The Physics of
Liquid Crystals}, (Oxford 1993).

\bibitem{remark2} Note that our anchoring energy depends on the
cosine and not on cosine squared, because we are dealing with a
ferronematic.

\bibitem{remark3} The periodicity is $2\pi$ because we deal with a
ferronematic where $\phi=0$ and $\phi=\pi$ are not equivalent.

\bibitem{c} One can obtain the solution to Eq. (\ref{diff_eq}) with
boundary conditions $\phi(x=0)=0$ and $\phi(x=\infty)=\Omega t$,
by taking $z(x,t)=\Omega-\partial\phi(x,t)/\partial t$. It then
follows that $z=\Omega[1-Erf(\frac12 x/\sqrt{Dt})]$. At the origin
we thus find that $\partial\phi(0)/\partial
x=\frac{2}{\sqrt{\pi}}\Omega \left(\frac{t}{D}\right)^{1/2}$.

\bibitem{ambeg} J. S. Langer and V. Ambegaokar, Phys. Rev. 
\textbf{164}, (1967) 498.




\end{thebibliography}
\end{document}